# Stacked conductive metal–organic framework nanorods for high-performance vacuum electronic devices

Zhengxin Guan [a], Jun Li [b, *], Han Wu [a], Xiaohong Chen [a], Wei Ou-Yang [a, *]

[a] Engineering Research Center for Nanophotonics & Advanced Instrument, Ministry of Education, School of Physics and Electronic Science, East China Normal University, 500 Dongchuan Road, Shanghai 200241, China.

[b] Department of Electronic Science and Technology, Tongji University, 4800 Caoan Road, Shanghai 201804, China.

* Corresponding authors: jun_li@tongji.edu.cn,  ouyangwei@phy.ecnu.edu.cn

## Abstract

Metal-organic frameworks (MOFs) possessing many unique features have been utilized in several fields in recent years. However, their application in field emission (FE) vacuum electronic device is hindered by their poor electrical conductivity. Herein, a novel conductive MOF of Cu-catecholate (Cu-CAT) with the nanorod length of 200 nm and conductivity of 0.01 S/cm is grown on the graphite paper (GP). Under an applied electric field, a large number of electrons can be emitted from the nanoscale emitter tips of MOF surface to the anode. The great field emission performance of Cu-CAT@GP cold cathode film including a low turn-on field of 0.59 V/μm and ultra-high field enhancement factor of 29622, even comparable to most carbon-based materials that have been widely investigated in FE studies, is achieved in this work. Meanwhile, Cu-CAT@GP film has a good electrical stability with a current attenuation of 9.4% in two hours. The findings reveal the cathode film fabricated by conductive MOF can be a promising candidate of cold electron source for vacuum electronic applications.

## 1. Introduction

Metal-organic frameworks (MOFs), consisting of metal ions and organic ligands, have attracted increasing attention owing to their porous structure, tunable pore size, numerous metal-containing active redox sites and high surface areas [1-4]. In view of their superior properties, many researches on MOFs are carried out in different fields, such as gas separation/desorption [5], sensing [6], energy storage [7, 8] and adsorption [9, 10]. However, most of the conventional MOFs are severely confined in some electrical or electrochemical applications because of their weak conductivity (<$10^{-6}$ S/cm) [11]. Their utilizations in electrode material applications have been limited to only serving as precursors/templates of porous carbons and metal oxides, or composites with other conductive materials such as carbon nanotubes (CNTs), graphene, conductive polymers, etc. [12-16] These strategies usually not only sacrifice the inherent large surface area and natural pores of MOFs, but also lead to complex synthesis. Thus, it is significant to fabricate novel MOFs which can be utilized as electrodes themselves and avoid complicated synthetic procedures. In recent years, several conductive MOFs have been investigated [17-19]. For example, the conductive MOF named Cu-catecholate (Cu-CAT) with outstanding conductivity (up to 0.2 S/cm) was first reported by Yaghi with his co-workers in 2012 [19]. Compared to conventional MOFs with poor conductivity, three electron transfer pathways can be envisaged in conductive MOFs, including through-space, through-bond and through-guest conduction [20].

A few conductive MOFs, as yet, have been directly used in oil/water separation, supercapacitors and lithium-ion batteries [21-23]. With their good conductivity and other splendid properties, field emission device (FED), a quantum tunnelling vacuum electronic device, is one of potential applications of the conductive MOFs [24, 25]. Up to now, FEDs have been investigated in several fields, such as field emission displays [25], microwave devices [26], ionic wind generators [27, 28] and triboelectric nanogenerators [29, 30]. Cold cathode, serving as an electron source, is the core of field emission devices. In the past few decades, many cathode materials such as metal molybdenum, metal oxides, carbon materials (carbon nanotubes, SiC and graphene), the MXene materials and so on have been explored and investigated [31-37]. Nevertheless, there is no research on the application of MOF materials in the field emission.

Herein, our experiment focused on the possibility of conductive MOF (Cu-CAT) application in the field of FE. We chose graphite paper (GP) as the substrate to grow Cu-CAT for the first time because of its high electrical conductivity, thermal conductivity and relatively low price comparing with substrates for Cu-CAT growth previously reported [38, 39]. As a consequence, a thin film of conductive Cu-CAT nanorods was grown on the GP and demonstrated excellent FE performance comparable with popular FE cathodes of carbon-based materials. The work provides a new direction for the selection of FE cathode materials and shows great potential for the development of FE devices.

## 2. Experimental section

The fabrication procedure of the conductive Cu-CAT nanorod cathode in this work is schematically shown in Fig. 1a. Similar to a typical procedure [40], copper acetate monohydrate (12 mg, 0.06 mmol) and HHTP (2,3,6,7,10,11-Hexahydroxytriphenylene) (9.8 mg, 0.03 mmol) were dispersed in 1 mL solvent mixture of deionized water/DMF (N, N-Dimethylformamide) (v: v = 1:1) under sonication for 15 min in a 10 ml teflon-lined stainless-steel autoclave. Note that in order that the reactants can be uniformly dispersed on the graphite paper serving as a substrate for growth of Cu-CAT nanorods, the solution continued to be sonicated for 10 minutes after the graphite paper was submerged into the reactant solution. The autoclave was then heated in an oven at the temperature of 85 °C for the crystallization to occur. After 24 h of reaction, the reactor was cooled to room temperature and the dark colored product grown on the graphite paper was achieved. The MOF nanorods coated graphite paper was then washed with deionized water for three times and subsequently treated by freeze-drying technique.

For the purpose of forming a complete cathode, it is also necessary to glue the coated graphite paper and fluorine-doped tin oxide (FTO) glass with double-sided conductive adhesive [41]. Next, the Cu-CAT cathode was assembled into a field emission device as one part of it. The inset in Fig. 4a shows the cross-section view of the FED. The FTO glass as anode was coated with a layer of phosphor to explore its luminescence properties. The distance between the cathode and anode of the FED was controlled at 990 μm.

## 3. Results and discussion

Fig. 1b shows the crystal structure of Cu-CAT. Each Cu ion coordinates with two adjacent 2,3,6,7,10,11-hexahydroxytriphenylene (HHTP) ligands to form a two-dimensional (2D) hexagonal lattice on the ab plane by virtue of honeycomb-like porous structure formed by π-stacking and π-conjugation. Effective orbital overlap between Cu ions and organic ligands makes Cu-CAT own good charge transfer characteristics [39]. The scanning electron microscopy (SEM) images in Fig. 1c-e illustrate the morphology of graphite paper and Cu-CAT nanorods grown on graphite paper (Cu-CAT@GP) respectively. Fig. 1c shows the graphite paper is composed of rolled graphite flakes, which contain numerous sharp edges in the microstructure. According to the Fig. 1d, it demonstrates that most MOF nanorods are tightly stacked together near the edge of the graphite sheet. Fig. S1a-c show the growth of Cu-CAT on graphite paper at different reaction times under the same hydrothermal conditions. The length of Cu-CAT nanorods can be controlled by reaction time. As shown in Fig. 1e, Cu-CAT can grow to

200 nm in length on graphite paper after hydrothermal reaction for 24 h. Note that the shape of Cu-CAT nanorods has been reported previously, while the Cu-CAT grown on graphite paper has a larger diameter and a shorter length [42]. Moreover, Cu, O and C elements are all detected from energy dispersive X-ray spectroscopy (EDS) elemental mapping images (Fig. S2a-c). The content ratio of O atom to Cu atom is 4.19:1 (Fig. S2d and Table S1), which is close to the Cu-CAT ($C_{36}H_{12}O_{12}Cu_3$) theoretical value ratio of 4:1, revealing that a thin film of Cu-CAT has been coated on graphite paper. The transmission electron microscope (TEM) image of the Fig. 2a demonstrates the stacked rod-like structure of Cu-CAT with a diameter of approximately 45 nm. In the Fig. 2b, a clear lattice strip with a spacing of 0.336 nm between Cu-CAT (002) crystal plane is observed.

Fig. 2c illustrates the X-ray diffraction (XRD) pattern of Cu-CAT nanorods. Note that the diffraction peak intensity of graphite paper in the region of 2θ between 25.5 and 27.5° is too strong to display. All diffraction peaks located at 4.7, 9.5, 12.6, 16.5, and 27.7° match well with the XRD patterns of Cu-CAT powder and simulated diffraction peaks, illustrating that the Cu-CAT nanorods grown on the graphite paper have similar crystal structure and crystallinity with the reported Cu-CAT [42]. The peak of (002) located at 27.7° corresponds to an interlayer spacing of 0.322 nm according to the well-known Bragg equation, which is similar to the value of (002) spacing analyzed from TEM image of Cu-CAT.

For the further investigation of specific surface chemical states of Cu-CAT nanorods, X-ray photoelectron spectroscopy (XPS) measurement was also performed. As shown in Fig. 2d, the main element peaks centered at binding energies (BEs) of 285.17, 532.7 and 935.08 eV are attributed to C1s, O1s, and Cu2p, respectively. The element peak centered at BE of 400.62 eV is attributed to N1s, which probably derives from a very small amount of residual DMF. The spectrum of C1s (Fig. S3) can be deconvoluted into four peaks at BEs of 284.4, 284.8, 285.5 and 287.1 eV, corresponding to C=C, C-C, C-O and C=O, respectively [43]. The O1s spectrum (Fig. 2e) can be fitted into three peaks at 531.6, 532.3 and 533.1eV, respectively assigned to O-Cu, O-C and O=C [44]. Fig. 2f shows the high-resolution Cu 2p spectrum of Cu-CAT nanowires. The peaks located at BE of 934.6 eV (Cu $2p_{3/2}$) and 954.6 eV (Cu $2p_{1/2}$), and the shake-up satellite (Sat.) peaks in 938.9-947.4 eV are the characteristic peaks of $Cu^{2+}$ [40]. The nitrogen sorption results and the corresponding pore size distribution of Cu-CAT (Fig. S4a, b) shows that the nanorods possess a large Brunauer–Emmett–Teller (BET) specific surface area of 229 $m^2/g$ and narrow pore size distribution of 0.9-1.2 nm.

The electronic property of Cu-CAT was featured by work function measurement using ultraviolet photoelectron spectroscopy (UPS). The work function is an important factor influencing FE performance [45]. Fig. 3a shows the UPS spectra of Cu-CAT nanorods having corrected bias. We find that the work function of synthesized Cu-CAT nanorods deposited on graphite paper is equal to 5.84 eV, determined from the secondary electron cut-off, similar to the reported values of 5.99 eV [46]. Such difference in work function could be attributed to the different synthesis environment that affects the surface state. The determination of the valence band (VB) edge is also an important parameter for band energy alignment diagram, as presented in Fig. 3b, VB edge value estimated of Cu-CAT is 1.79 eV. The energy diagram in Fig. 3c schematizes the work function and ionization energy reduction of Cu-CAT.

The electrical conductivity can also impact the FE performance. To some extent, the conductivity represents the ability of the cathode material to transfer electrons to the field emitters and generate electron beams. A cold cathode with higher electrical conductivity tends to have better FE performance. In our previous work, the field emission performance of SiC nanowire cathode was enhanced by adding conductive polymers to increase the conductivity of SiC [47]. The electrical conductivity of Cu-CAT nanorods in this work is measured to be 0.01 S/cm, similar to the conductivity of Cu-CAT reported before [48]. As a part of cathode film, graphite paper (measured to be

2564 S/cm) with high conductivity can quickly transport carriers to emission tips of Cu-CAT nanorods and emit electron beams under external voltage, so as to obtain excellent field emission performance.

To explore the field emission performance, the current density versus applied electric field (*J*−*E*) plot for field emission characteristics of the Cu-CAT nanorods cathode is shown in Fig. 4a. Turn-on field ($E_{on}$, defined at a current density of 10 µA/cm$^2$) and threshold field ($E_{th}$, defined at a current density of 100 µA/cm$^2$) are 0.59 and 0.81 V/µm, respectively. These values are best among the state-of-the-art reported devices [36,41,49]. As is known, field emission cathode materials often have protrusions in the microscopic morphology, which make them play a role in greatly enhancing the local electric field in field emission. The relationship between the local electric field ($E_{local}$) and the applied electric field (*E*) can be described as $E_{local} = \beta E$, where *β*, the field enhancement factor, is also one of important factors for judging field emission performance. It can be calculated by Fowler−Nordheim (F−N) theory [50] which is described by

$$J = A\Phi^{-1}\beta^2 e^2 \exp\left(-\frac{B\Phi^{\frac{3}{2}}}{\beta E}\right) \ , \tag{1}$$

where both A and B are constants, corresponding to 1.54 × 10$^{-6}$ A·eV·V$^{-2}$ and 6.83 × 10$^3$ V·eV$^{3/2}$·µm$^{-1}$, respectively, and *ϕ* is the work function of the cathode (5.84 eV in this study). The F−N plot of Cu-CAT nanorods (Fig. S5) displays the value of $\beta$ can be calculated as 29622 according to the slope of the plot.

As shown in Fig. 4b and Table S2, compared with several field emission performances of newly reported cathode materials, Cu-CAT surpasses most of its counterparts in the list. It can be even comparable to some carbon-based materials such as CNTs and SiC, which are always considered as popular candidates for field emission cathodes.

In addition, we also studied the device stability of Cu-CAT nanorod cathode. The field emission stability has been shown in Fig. 5a and the original current density was set to 100 μA/cm$^2$. According to the plot where the normalized current density changes with time, the current density has no obvious degradation within 120 minutes. The current density decreased from 100 to 90.6 μA/cm$^2$, with an attenuation of 9.4%. Moreover, the illustrations of insets in Fig. 5a show the luminous uniformity of sample and the current fluctuation range concentrated within ±20% which demonstrates the MOF-FED has a pretty good stability in vacuum compared with recent reports [34,51]. Meanwhile, no obvious change of Cu–CAT@GP's XRD pattern can be found before and after field emission test, indicating the excellent stability of the MOF cathode in FED under vacuum. This can be partly attributed to graphite paper. As the substrate for growing Cu-CAT nanorods and one part of the cathode material film, the graphite paper with good thermal conductivity can make the field emission cathode weaken the influence of thermal effect caused by high current density reducing the damage to the morphology and microstructure, and thus improve the stability.

A comparison of the feasibility of previously reported different substates used to grow Cu-CAT and the graphite paper used in this work is shown in Table S3, indicating that graphite paper has great advantages in terms of electrical conductivity, thermal conductivity, and price.

## 4. Conclusions

In summary, we fabricated stacked Cu-CAT nanorods grown on graphite paper from bottom to top through a hydrothermal synthesis method. Compared with other common substrates for growing Cu-CAT nanorods, the graphite paper used in this study is not only relatively inexpensive but also has good electrical and thermal conductivity. The Cu-CAT film grown on graphite paper (Cu-CAT@GP film) has a large specific surface area (229 $m^2/g$), small pore size (0.9-1.2 nm) and high electrical conductivity (0.01 S/cm). Moreover, outstanding field emission performance of the conductive MOFs is proved. Because of their rod-like one-dimensional (1D) structure and good conductivity, Cu-CAT nanorods possess splendid FE properties. Their turn-on and threshold applied fields can be 0.59 and 0.81 V/μm, respectively. Simultaneously, it also owns high field enhancement factor of 29622 and great current stability. These findings reveal that conductive metal-organic framework can be also a promising candidate for vacuum electronic devices.

## Conflicts of interest

There are no conflicts to declare.

## Acknowledgements

This work was financially supported by National Natural Science Foundation of China (Grant No. 61771198) and Natural Science Foundation of Shanghai (Grant No. 17ZR1447000). We acknowledge Prof. S. Liu and Prof. M. Hu at East China Normal University, Shanghai, China for their help in material preparation and characterization.

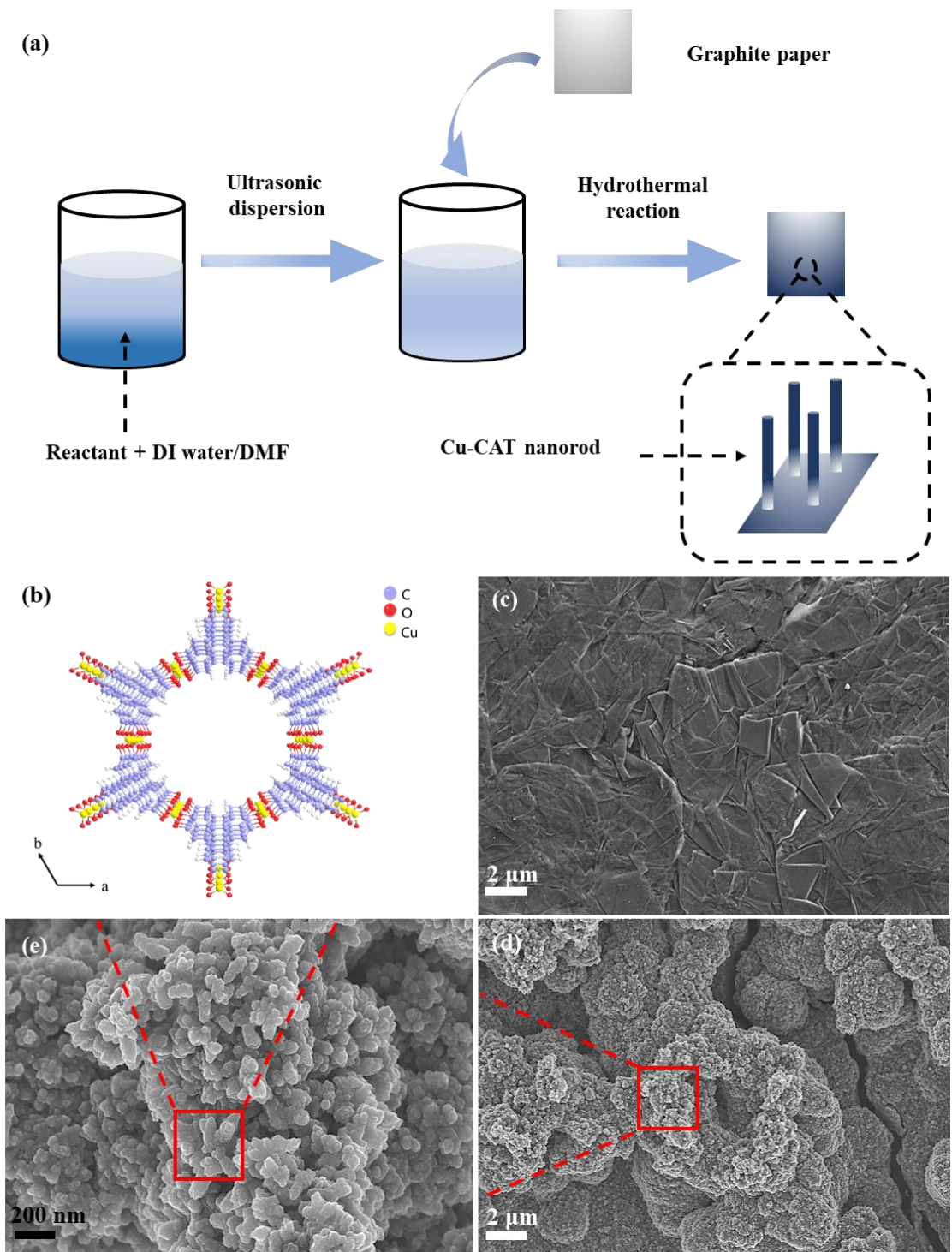

Fig. 1 (a) Preparation illustration of Cu-CAT@GP. (b) The simulated crystal structure of Cu-CAT along the c-axis. (c) SEM of the graphite paper. (d and e) SEM of rough Cu-CAT nanorods grown on graphite paper.

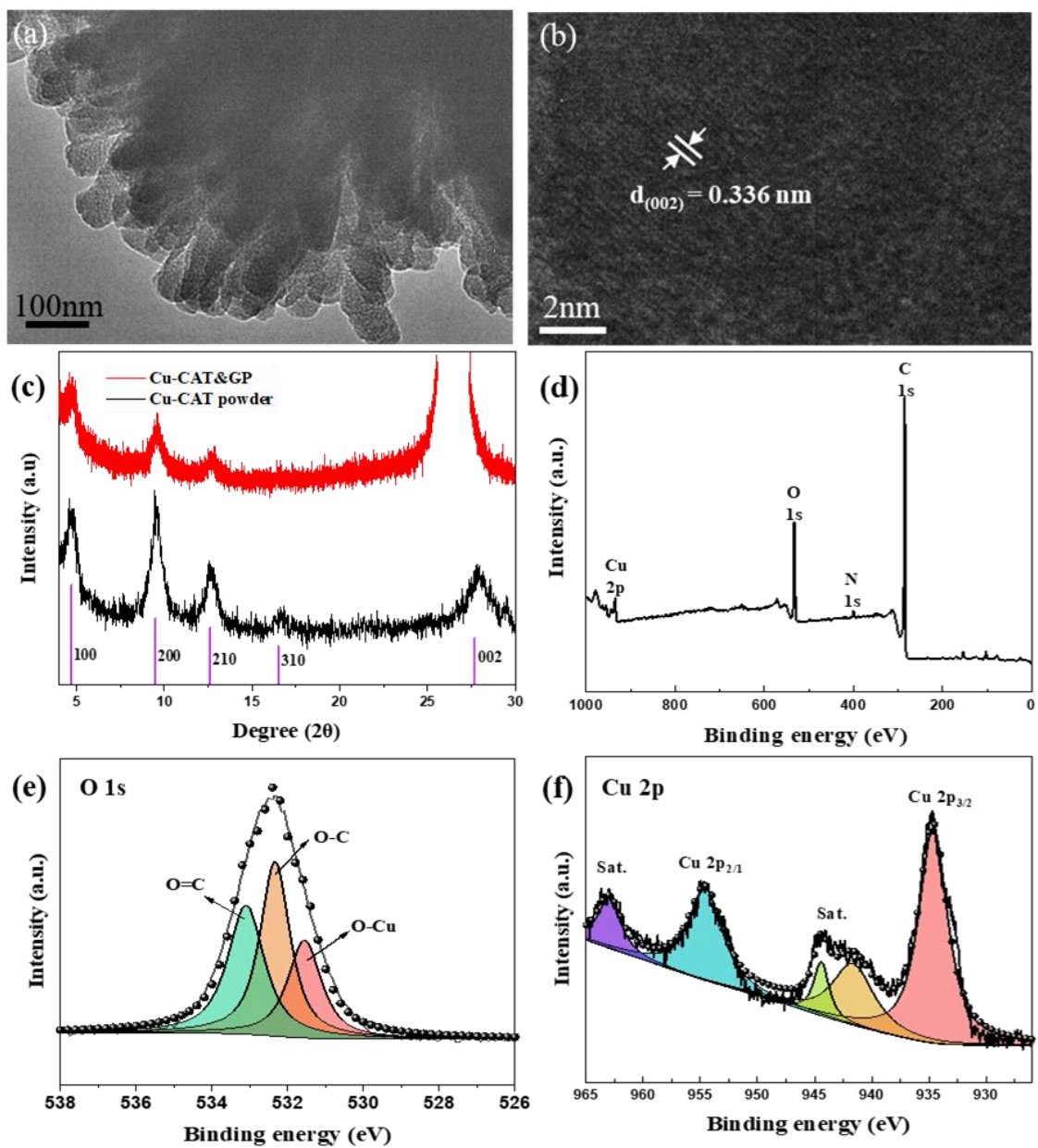

Fig. 2 (a and b) TEM images of Cu-CAT rough nanorods. (c) XRD patterns for Cu–CAT. (d) XPS survey sepctrum of Cu-CAT. (e and f) High-resolution XPS spectra of O 1s and Cu 2p, respectively.

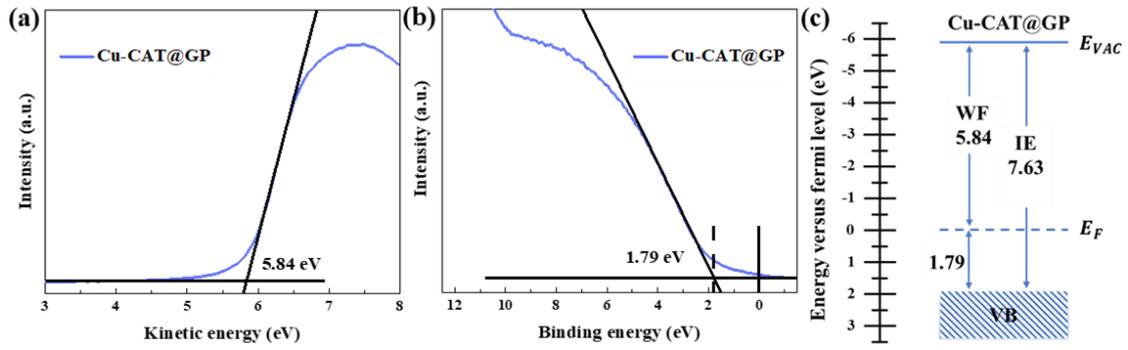

Fig. 3 (a) UPS spectrum around the secondary electron cut-off of Cu-CAT. (b) UPS spectrum in the valence band (VB) region of Cu-CAT. (c) Energy scheme for Cu-CAT with respect to the $E_F$. IE, ionization energy. $E_F$, Fermi level. $E_{Vac}$, vacuum level. WF, work function.

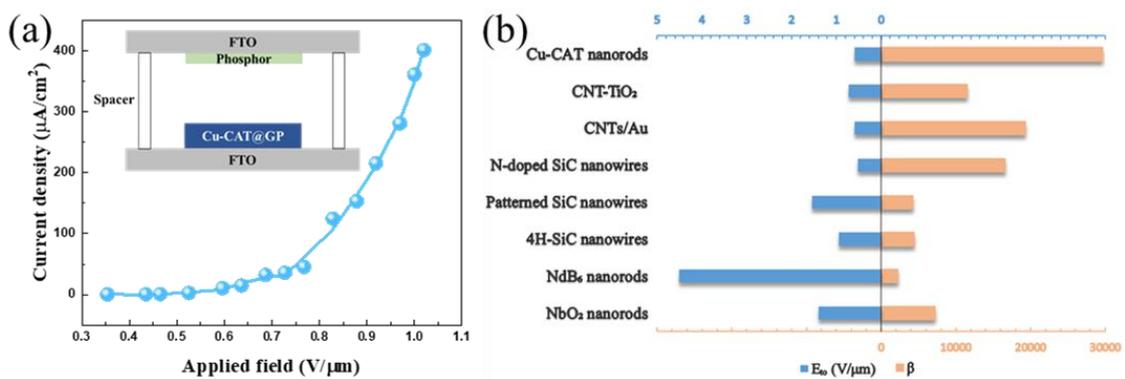

Fig. 4 (a) The typical *J–E* curve of Cu-CAT@GP and cross-section view of the FED (inset in a). (b) Field emission performance comparison of Cu–CAT nanorods and recent reported 1D nanostructured emitters.

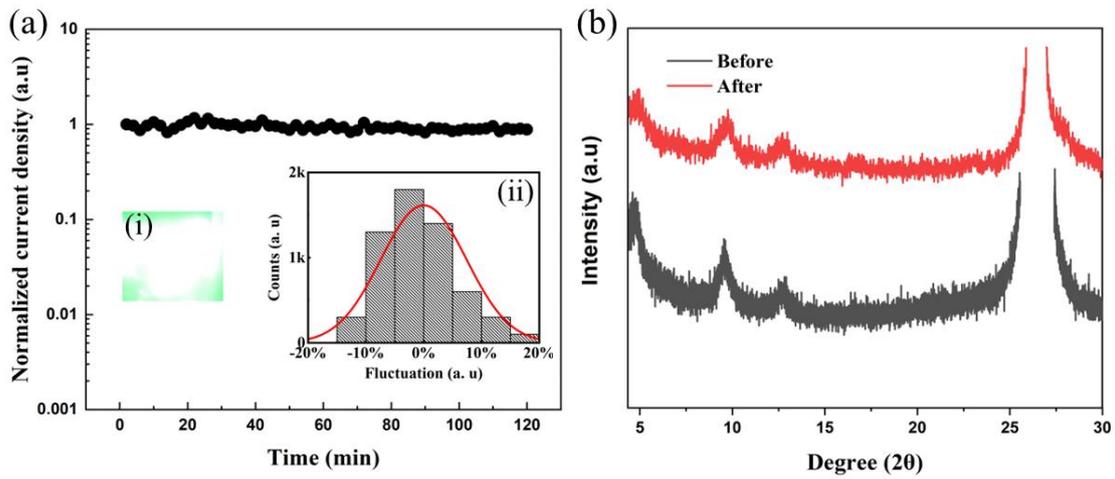

Fig. 5 (a) Field emission stability of Cu-CAT@GP cathode; inset (i) is luminance picture of FED with Cu-CAT@GP cathode; inset (ii) is frequency histogram and normalized curve of current density fluctuation. (b) XRD patterns for the Cu–CAT@GP cathode before and after field emission test.

**Highlights**

- A conductive MOF film composed of stacked Cu-CAT nanorods with conductivity of 0.01 S/cm is fabricated on graphite paper.
- Field emission performance of MOF materials is first demonstrated for the conductive Cu-CAT film.
- The FED presents a pretty low turn-on field of 0.59 V/μm, comparable with the state-of-the-art CNT-based cathode materials.

**Graphical abstract**

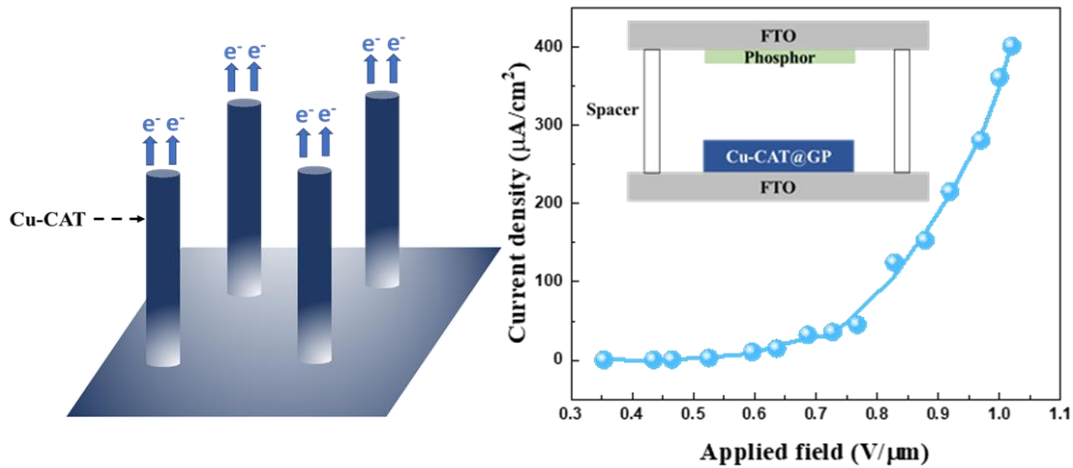